\documentclass[aps,prl,preprint,floats,epsfig,showpacs,tightenlines,superscriptaddress,byrevtex]{revtex4}  

\usepackage{graphicx} 
\usepackage{dcolumn}  

\graphicspath{{ps}}

\def\Journal#1#2#3#4{{#1} {\bf #2}, #3 (#4)}

\def\NIMA{{\em Nucl. Instrum. Meth.} A}
\def\NPB{{\em Nucl. Phys.} B}
\def\PLB{{\em Phys. Lett.}  B}
\def\PRL{\em Phys. Rev. Lett.}
\def\PRD{{\em Phys. Rev.} D}
\def\ZPC{{\em Z. Phys.} C}

\def\Journal#1#2#3#4{{#1} {\bf #2}, #3 (#4)}

\def\NIMA{{\em Nucl. Instrum. Meth.} A}
\def\NPB{{\em Nucl. Phys.} B}
\def\PLB{{\em Phys. Lett.}  B}
\def\PRL{\em Phys. Rev. Lett.}
\def\PRD{{\em Phys. Rev.} D}
\def\ZPC{{\em Z. Phys.} C}

\def\EPJC{{\em Eur. Phys. J.} C}
\def\IJMPA{{\em Int. J. Mod. Phys.} A}

\def\GeV{{\rm~GeV}}

\def\PM#1#2{\,^{+#1}_{-#2}}

\begin{document}


\preprint{\vbox{\hbox{\hfil BELLE-CONF-0415}
}}


\title{ \quad\\[1cm] 
Measurement of the Differential $q^2$ Spectrum and Forward-Backward
Asymmetry for $B \to K^{(*)} \ell^+ \ell^-$
}



\affiliation{Aomori University, Aomori}
\affiliation{Budker Institute of Nuclear Physics, Novosibirsk}
\affiliation{Chiba University, Chiba}
\affiliation{Chonnam National University, Kwangju}
\affiliation{Chuo University, Tokyo}
\affiliation{University of Cincinnati, Cincinnati, Ohio 45221}
\affiliation{University of Frankfurt, Frankfurt}
\affiliation{Gyeongsang National University, Chinju}
\affiliation{University of Hawaii, Honolulu, Hawaii 96822}
\affiliation{High Energy Accelerator Research Organization (KEK), Tsukuba}
\affiliation{Hiroshima Institute of Technology, Hiroshima}
\affiliation{Institute of High Energy Physics, Chinese Academy of Sciences, Beijing}
\affiliation{Institute of High Energy Physics, Vienna}
\affiliation{Institute for Theoretical and Experimental Physics, Moscow}
\affiliation{J. Stefan Institute, Ljubljana}
\affiliation{Kanagawa University, Yokohama}
\affiliation{Korea University, Seoul}
\affiliation{Kyoto University, Kyoto}
\affiliation{Kyungpook National University, Taegu}
\affiliation{Swiss Federal Institute of Technology of Lausanne, EPFL, Lausanne}
\affiliation{University of Ljubljana, Ljubljana}
\affiliation{University of Maribor, Maribor}
\affiliation{University of Melbourne, Victoria}
\affiliation{Nagoya University, Nagoya}
\affiliation{Nara Women's University, Nara}
\affiliation{National Central University, Chung-li}
\affiliation{National Kaohsiung Normal University, Kaohsiung}
\affiliation{National United University, Miao Li}
\affiliation{Department of Physics, National Taiwan University, Taipei}
\affiliation{H. Niewodniczanski Institute of Nuclear Physics, Krakow}
\affiliation{Nihon Dental College, Niigata}
\affiliation{Niigata University, Niigata}
\affiliation{Osaka City University, Osaka}
\affiliation{Osaka University, Osaka}
\affiliation{Panjab University, Chandigarh}
\affiliation{Peking University, Beijing}
\affiliation{Princeton University, Princeton, New Jersey 08545}
\affiliation{RIKEN BNL Research Center, Upton, New York 11973}
\affiliation{Saga University, Saga}
\affiliation{University of Science and Technology of China, Hefei}
\affiliation{Seoul National University, Seoul}
\affiliation{Sungkyunkwan University, Suwon}
\affiliation{University of Sydney, Sydney NSW}
\affiliation{Tata Institute of Fundamental Research, Bombay}
\affiliation{Toho University, Funabashi}
\affiliation{Tohoku Gakuin University, Tagajo}
\affiliation{Tohoku University, Sendai}
\affiliation{Department of Physics, University of Tokyo, Tokyo}
\affiliation{Tokyo Institute of Technology, Tokyo}
\affiliation{Tokyo Metropolitan University, Tokyo}
\affiliation{Tokyo University of Agriculture and Technology, Tokyo}
\affiliation{Toyama National College of Maritime Technology, Toyama}
\affiliation{University of Tsukuba, Tsukuba}
\affiliation{Utkal University, Bhubaneswer}
\affiliation{Virginia Polytechnic Institute and State University, Blacksburg, Virginia 24061}
\affiliation{Yonsei University, Seoul}
  \author{K.~Abe}\affiliation{High Energy Accelerator Research Organization (KEK), Tsukuba} 
  \author{K.~Abe}\affiliation{Tohoku Gakuin University, Tagajo} 
  \author{N.~Abe}\affiliation{Tokyo Institute of Technology, Tokyo} 
  \author{I.~Adachi}\affiliation{High Energy Accelerator Research Organization (KEK), Tsukuba} 
  \author{H.~Aihara}\affiliation{Department of Physics, University of Tokyo, Tokyo} 
  \author{M.~Akatsu}\affiliation{Nagoya University, Nagoya} 
  \author{Y.~Asano}\affiliation{University of Tsukuba, Tsukuba} 
  \author{T.~Aso}\affiliation{Toyama National College of Maritime Technology, Toyama} 
  \author{V.~Aulchenko}\affiliation{Budker Institute of Nuclear Physics, Novosibirsk} 
  \author{T.~Aushev}\affiliation{Institute for Theoretical and Experimental Physics, Moscow} 
% ...
\collaboration{The Belle Collaboration}

\noaffiliation

\begin{abstract}

We report measurements of the differential $q^2$ spectrum and
the forward-backward asymmetry for $B \to K^{(*)} \ell^+\ell^-$,
where $\ell$ represents an electron or a muon,
with a data sample of 253 fb$^{-1}$ accumulated
on the $\Upsilon(4S)$ resonance with the Belle detector at KEKB.
We also present measurements of the branching
fractions and their ratios.
\end{abstract}

\pacs{13.20.He, 11.30.Hv }  

\maketitle


%
%
%

{\renewcommand{\thefootnote}{\fnsymbol{footnote}}}


\newpage

\normalsize

\setcounter{footnote}{0}
\newpage

\normalsize

%
%


Flavor-changing neutral current (FCNC) processes are forbidden at
tree level in the Standard Model (SM); rather, they proceed at a low
rate via loop or box diagrams. If additional like diagrams with
non-SM particles contribute to such a decay, their amplitudes will
interfere with the SM amplitudes and thereby modify the decay rate as
well as other properties.
This feature makes FCNC
processes an ideal place to search for new physics. 

Measurements of the radiative penguin decay $B\to
X_{s}\gamma$~\cite{belle-btosgamma,cleo-btosgamma,aleph-btosgamma},
which are consistent with the SM prediction, strongly constrain the
magnitude---but not the sign---of the effective Wilson coefficient $C_7$.
This is an important limitation, since non-SM contributions can change
the sign of $C_7$ without changing the $B\to
X_{s} \gamma$ branching fraction~\cite{C7}.

The $b \to s \ell^{+} \ell^{-}$ process is promising from this point
of view, since not only the photonic penguin diagram but also the
$Z$-penguin and box diagrams contribute to this decay mode. As a
result, the Wilson coefficients $C_7$, $C_9$
and $C_{10}$ can be completely determined. The first observations of $B\to K
\ell^+\ell^-$, $B\to K^* \ell^+\ell^-$ and inclusive $B \to X_{s} \ell^+
\ell^-$ decays were reported by the Belle
Collaboration~\cite{BelleKll,BelleKstarll,BelleXsll}.
The measured branching fractions of these decay modes were used to
exclude a large area of the allowed region in the $C_9$-$C_{10}$ plane~\cite{ALGH,Lunghi}.
However, the determination of the sign of $C_7$ (as well as of $C_9$ 
and $C_{10}$)
requires precise measurements of the distribution in squared dilepton
momentum $q^2$ and the forward-backward
asymmetry in these decay modes~\cite{C7C9C10}. 

In this paper, we present preliminary results of improved measurements
of $B \to K \ell^+ \ell^-$ and $B \to
K^* \ell^+ \ell^-$ using data produced in $e^+{}e^{-}$
annihilation at the KEKB asymmetric collider \cite{kekb} and
collected with the Belle detector. 
The data sample corresponds to
253~fb${}^{-1}$ taken at the $\Upsilon(4S)$ resonance and contains
approximately 275 million $B\overline{B}$ pairs.

The Belle detector is a large-solid-angle magnetic
spectrometer that consists of a silicon vertex detector (SVD),
a 50-layer central drift chamber (CDC), an array of
aerogel threshold \v{C}erenkov counters (ACC),
a barrel-like arrangement of time-of-flight
scintillation counters (TOF), and an electromagnetic calorimeter (ECL)
comprised of CsI(Tl) crystals located inside
a super-conducting solenoid coil that provides a 1.5~T
magnetic field.  An iron flux-return located outside of
the coil is instrumented to detect $K_L^0$ mesons and to identify
muons (KLM).  The detector is described in detail elsewhere~\cite{NIM}.
The data were collected with two different inner detector configurations.
For the first sample
of 152 million $B\bar{B}$ pairs, a 2.0 cm radius beampipe
and a 3-layer silicon vertex detector were used;
for the latter 123 million $B\bar{B}$ pairs,
a 1.5 cm radius beampipe, a 4-layer silicon detector,
and a small-cell inner drift chamber were used~\cite{Ushiroda}.

In this analysis, primary charged tracks---except for the daughters
from $K_S \to \pi^+ \pi^-$ decays---are required to have impact
parameters relative to the interaction point of less than 5.0~cm along
the $z$ axis (aligned opposite the positron beam) and less than 0.5~cm in the $r\phi$ plane that is transverse to this axis.  This requirement reduces the combinatorial background from photon conversion, beam-gas and beam-wall events. 

Charged tracks are identified as either kaons or pions by a likelihood ratio
based on the CDC specific ionization, time-of-flight information and the
light yield in the ACC.  This classification is superseded for a track that is
identified as an electron or for a pion-like track that is identified as a muon.
Electrons are identified from the ratio of shower energy of the matching ECL cluster to
the momentum measured by the CDC, the transverse shower shape of this cluster, the
specific ionization in the CDC and the ACC response. We require
that the electron momentum be greater than $0.4$~GeV/$c$ to reach the ECL.   
Muons are identified by their penetration depth and transverse scattering
in the KLM. The muon momentum is required to exceed 0.7~GeV/$c$.  The muon
identification criteria are more stringent for momenta below 1.0~GeV/$c$ to suppress
misidentified hadrons.

Photons are selected from isolated neutral clusters in the ECL 
with energy greater than $50$~MeV and a shape 
that is consistent with an electromagnetic shower. 
Neutral pion candidates are reconstructed from pairs of
photons, and are required to have an invariant mass 
within $10$~MeV/$c^2$ of the nominal $\pi^{0}$ mass and
a laboratory momentum greater than 0.1 GeV/$c$.
The pion momentum is recalculated by constraining the invariant mass
to the nominal $\pi^0$ mass.
$K^{0}_{S}$ candidates are reconstructed from oppositely 
charged pions that have invariant masses within $15$~MeV/$c^{2}$ 
of the nominal $K^{0}_{S}$ mass. We impose additional criteria based on
the radial impact parameters of the pions ($\delta r$), 
the distance between the closest approaches of the pions 
along the beam direction ($\delta z$),
the distance of the vertex from the interaction point ($l$), 
and the azimuthal angle difference between the
vertex direction and the $K^{0}_{S}$ momentum direction ($\delta\phi$).
These variables are combined as follows: for $p_{K^{0}_{S}} < 0.5$~GeV/$c$,
$\delta z < 8$~mm, $\delta r > 0.5$~mm, and $\delta\phi <0.3$~rad are required;
for  $0.5$~GeV/$c < p_{K^{0}_{S}} < 1.5$~GeV/$c$, $\delta z < 18$~mm, 
$\delta r > 0.3$~mm, $\delta\phi <0.1$~rad, and $l > 0.8$~mm are required; 
and for  $p_{K^{0}_{S}} > 1.5$~GeV/$c$, $\delta z < 24$~mm, 
$\delta r > 0.2$~mm, $\delta\phi <0.03$~rad, and $l > 2.2$~mm are required.

$K^{*}$ candidates are formed by combining 
a kaon and a pion: $K^{+}\pi^{-}$, 
$K^{0}_{S}\pi^{+}$ or $K^{+}\pi^{0}$~\cite{cc}.
The $K^{*}$ invariant mass is required to lie within 
$75$~MeV/$c^2$ of the nominal $K^{*}$ mass. 
For modes involving neutral pions, 
combinatorial backgrounds
are reduced by the additional requirement
$\cos\theta_{\rm{hel}}<0.8$,
where $\theta_{\rm{hel}}$
is defined as the angle between the opposite of $B$ momentum
and the kaon momentum direction in the $K^{*}$ rest frame.

$B$ candidates are reconstructed from a $K^{(*)}$ candidate and an
oppositely-charged lepton pair. We use two variables defined in the CM frame
to select $B$
candidates: the beam-energy constrained mass $M_{\rm{bc}} = \sqrt{ {E_{\rm{beam}}^*}^2 -
{p_{B}^*}^{2}}$ and the energy difference
$\Delta E = E_{B}^* - E_{\rm{beam}}^* $, where 
$p_{B}^*$ and $E_B^*$ are the measured momentum and energy, respectively, of the $B$
candidate, and $E_{\rm{beam}}^*$ is the beam energy~\cite{aster}. When multiple candidates are
found in an event, we select the candidate with the smallest value of $|\Delta E|$.

Backgrounds from $B \to J/\psi(\psi') K^{(*)}$ are rejected using the
dilepton invariant mass.
The veto windows are defined as\\
\hspace*{2cm}
\begin{tabular}{r@{}l@{\qquad}l}
  $-0.25~\textrm{GeV/}c^{2}$ & ${}< M_{e^{+}e^{-}}     - M_{J/\psi}   < 0.07~\textrm{GeV/}c^{2}$ & for $K^{*}$ modes \\
  $-0.20~\textrm{GeV/}c^{2}$ & ${}< M_{e^{+}e^{-}}     - M_{J/\psi}   < 0.07~\textrm{GeV/}c^{2}$ & for $K$ modes \\
  $-0.20~\textrm{GeV/}c^{2}$ & ${}< M_{e^{+}e^{-}}     - M_{\psi'} < 0.07~\textrm{GeV/}c^{2}$ & for $K^{*}$ and $K$ modes\\
  $-0.15~\textrm{GeV/}c^{2}$ & ${}< M_{\mu^{+}\mu^{-}} - M_{J/\psi}   < 0.08~\textrm{GeV/}c^{2}$ & for $K^{*}$ modes\\
  $-0.10~\textrm{GeV/}c^{2}$ & ${}< M_{\mu^{+}\mu^{-}} - M_{J/\psi}   < 0.08~\textrm{GeV/}c^{2}$ & for $K$ modes\\
  $-0.10~\textrm{GeV/}c^{2}$ & ${}< M_{\mu^{+}\mu^{-}} - M_{\psi'} < 0.08~\textrm{GeV/}c^{2}$ & for $K^{*}$ and $K$ modes
\end{tabular}\\
If a photon with energy less than
500~MeV is found in the 50~mrad cone along the electron momentum
direction, we reapply the above vetoes with the invariant mass calculated including this
photon to reject the background with a bremsstrahlung photon, $J/\psi
(\psi') \to e^+e^-\gamma(\gamma)$.
For $K^* e^+ e^-$ modes, $B \to J/\psi \, K$ can be a background
if a bremsstrahlung photon is missed and a pion from the other $B$
meson in the event is included.  We suppress this background by the following
prescription: the included pion is discarded and an unobserved bremsstrahlung
photon is added with a direction parallel to the electron or positron and
an energy that gives $\Delta E = 0$ for the $B$ candidate.
If the dilepton mass and the beam-energy constrained mass are consistent
with a $B \to J/\psi \, K$ event, the candidate is vetoed.

We suppress background from photon conversions and $\pi^{0}$ Dalitz
decays by requiring the dielectron mass to satisfy $M_{e^{+}e^{-}} >
0.14~$GeV/$c^{2}$. This cut eliminates possible peaking background from
$B \to K^* \gamma$ and $B \to K^{(*)} \pi^0$.

Background from continuum $q\overline{q}$ events is suppressed using
event topology. Continuum events have a jet-like shape while
$B\overline{B}$ events have a spherical shape in the center-of-mass
frame. 
A Fisher discriminant $\cal{F}$~\cite{fd} is calculated from the energy
flow in 9 cones along the $B$ candidate sphericity axis and the
normalized second Fox-Wolfram moment $R_{2}$~\cite{fw}.  
We combine this with the cosine of the polar angle $\cos\theta_B^*$
of the $B$ meson flight direction 
and the cosine of the polar angle $\cos\theta^{*}_{\rm sph}$ of the $B$ meson sphericity axis to define likelihoods $\cal{L}_{\rm{sig}}$ and
$\cal{L}_{\rm{cont}}$ for signal and continuum background, respectively,
and then cut on the likelihood ratio $\cal{R}_{\rm{cont}} =
\cal{L}_{\rm{sig}} / (\cal{L}_{\rm{sig}} + \cal{L}_{\rm{cont}})$.
For the muon mode, $|\cos\theta^{*}_{\rm{sph}}|$ is not used since its
distribution is nearly the same for signal and continuum within the
detector acceptance.

The dominant background from $B\overline{B}$ events is due to
semileptonic $B$ decays. The missing energy of the event,
$E_{\mathrm{miss}} = 2 E_{\rm beam}^* - E_{\rm vis}^*$ 
where $E_{\rm vis}^*$ is a total visible energy in the event,
is used to suppress this background since 
the undetected neutrinos carry away a substantial amount of energy.
The $B$ meson flight angle, $\cos\theta_B^*$, is also used to suppress combinatorial
background in $B\overline{B}$ events.  
We combine $E_{\mathrm{miss}}$ and $\cos\theta^{*}_{B}$ into signal and
$B\overline{B}$-background likelihoods and cut on the
likelihood ratio ${\cal{R}}_{B\overline{B}}$, defined similarly to
${\cal{R_{\rm{cont}}}}$.

The $B \to K^{(*)} \ell^+ \ell^-$ decays are generated according to
Ref~\cite{ALGH} and then processed by a GEANT-based 
Monte Carlo~(MC) simulation to estimate the efficiencies.
The signal box is defined as $|M_{\rm bc}-M_{B}|<0.007$~GeV/$c^{2}$ for
both lepton modes and
$-0.055$~GeV$<\Delta E<0.035$~GeV ($|\Delta E|<0.035$~GeV) for the electron
(muon) mode.
We make distinct selections on ${\cal{R}}_{\rm cont}$ and ${\cal{R}}_{B\overline{B}}$
for each decay mode.
The detection efficiencies are estimated from the MC samples and are summarized in Table \ref{tab:results}.

\begin{figure}
\begin{center}
  \includegraphics[scale=0.9]{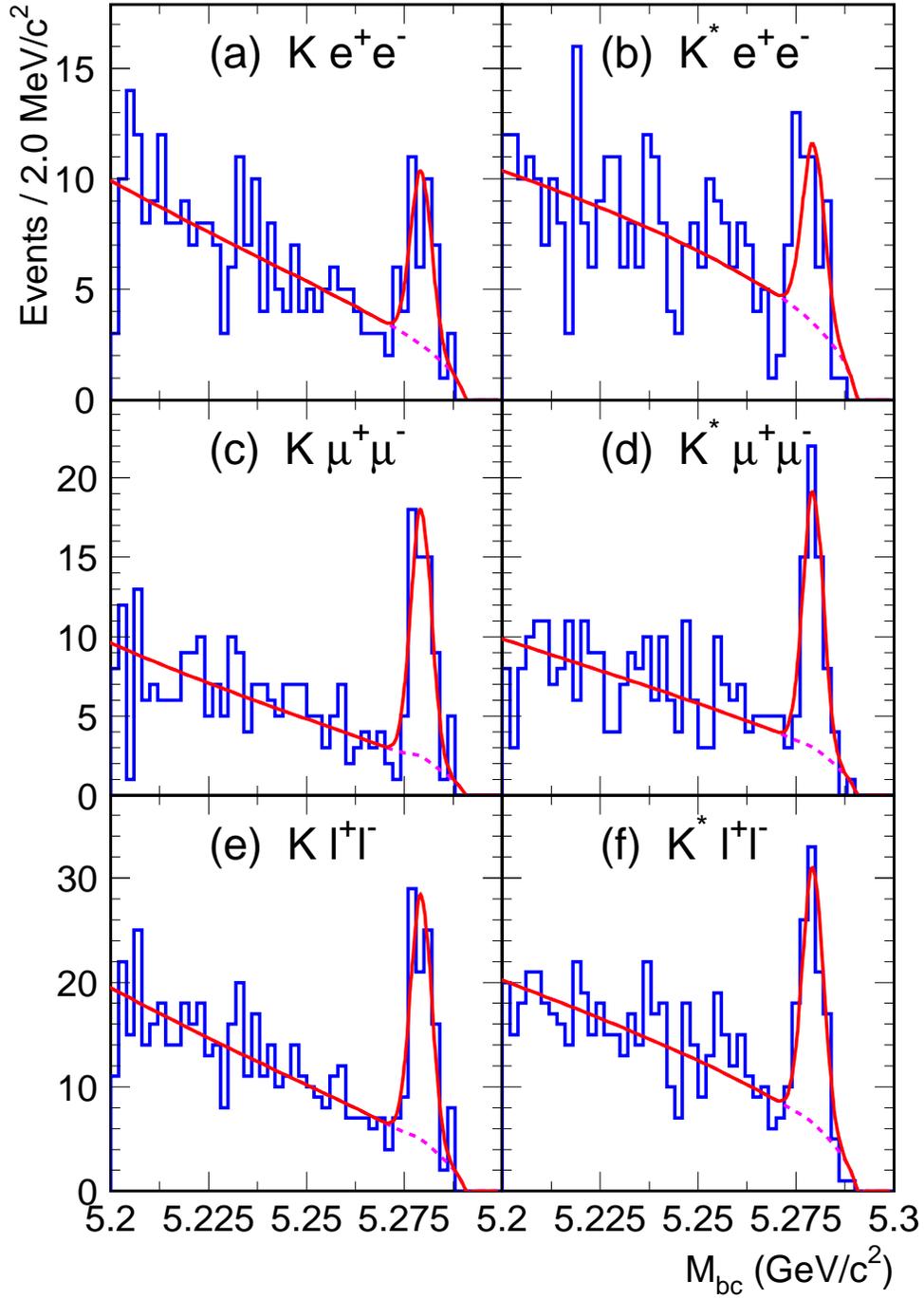}
\end{center}
\caption{$M_{\rm bc}$ distributions for (a) $B\to K e^{+} e^{-}$, (b)
$B \to K^{*} e^{+} e^{-}$, (c) $B\to K \mu^{+} \mu^{-}$, (d) $B \to
K^{*} \mu^{+} \mu^{-}$, (e) $B\to K \ell^{+} \ell^{-}$ and (f) $B \to
K^{*} \ell^{+} \ell^{-}$ samples. The solid
and dashed curves are the fit results of the total and background
contributions.
}
\label{fig:mbcfit}
\end{figure}

To determine the signal yield, we perform a binned maximum-likelihood fit to each $M_{\mathrm{bc}}$ distribution.
The expected number of signal events  is 
calculated as a function of $M_{\mathrm{bc}}$ using a Gaussian  signal distribution plus  background functions.
The mean and the width of the signal Gaussian 
are determined using observed $J/\psi K^{(*)}$ events.
A MC study shows that the width has no dependence  on the dilepton invariant mass.
We consider backgrounds from three sources: semileptonic decays,
decays containing $J/\psi$ and $\psi'$ mesons, and double misidentification of hadrons as leptons.
The background from semileptonic decays is parameterized by 
the ARGUS function~\cite{argus}.
The shape is determined from large $B\bar{B}$ and continuum MC samples,
each containing at least one oppositely charged lepton pair.
The shape parameter obtained from MC is consistent with that
taken from a data sample of $B \to K^{(*)}e^{\pm}\mu^{\mp}$ candidates.
The residual background from $J/\psi$ and $\psi'$ mesons that cannot
be removed by the $\psi^{(')}$ veto windows is estimated from a large MC sample of
$J/\psi$ and $\psi'$ inclusive events and parameterized by an ARGUS
function and a Gaussian. 
The background contribution due to misidentification of hadrons 
as leptons is  parameterized
by another ARGUS function and a Gaussian.
The ARGUS function represents the combinatorial background while
the Gaussian represents the component that forms a peak
in the $M_{\rm bc}$ distribution. 
The shape and normalization of this background are 
fixed using the $B\to K^{(*)}h^+h^-$ data sample (where $h$ refers to a pion or kaon).
All $K^{(*)} h^{+} h^{-}$ combinations are weighted by the momentum-
and polar angle-dependent probability of misidentifying $K^{(*)} h^+h^-$ as $K^{(*)} \ell^+\ell^-$.
This study yields $1.06 \pm 0.09$ $K h^+h^-$ events and $0.60 \pm
0.05$ $K^* h^+h^-$ events in the peak region. 
Other backgrounds with misidentified leptons are negligible.
The normalizations of the signal and the background from real leptons are floated in the fit. 
The fit results are shown in Fig.~\ref{fig:mbcfit} and Table~\ref{tab:results}.

We observe $78.5^{+10.7}_{-10.0}$ $B \to K^{*} \ell^{+} \ell^{-}$ signal
with a significance of $11.0$ and $82.2^{+11.4}_{-10.7}$ $B \to K \ell^{+} \ell^{-}$
signal with a significance of $10.1$.
The significance is defined as 
$\sqrt{-2\ln({\cal L}_0/{\cal L}_{\rm{max}})}$, 
where ${\cal L}_{\rm{max}}$ is the maximum likelihood 
in the $M_{\mathrm{bc}}$ fit and ${\cal L}_0$ is the likelihood 
of the best fit when the signal yield is constrained to be zero. Since the signal
and background shapes are fixed in the fitting, the error due to
uncertainty in shape parameters is included in the systematic error. 
We determine this by varying each shape parameter by $\pm1\sigma$ and
recalculating the signal yield.  We quote the smallest resulting
significance in our determinations above.

\begin{table}[htbp]
 \caption{Summary of the fit results and branching fractions. 
  Signal yield estimated from the $M_{\mathrm{bc}}$ fit,
  detection efficiency for each mode, obtained branching fraction, 
  90\% confidence level upper limit of the branching fraction
  and the significance of the signal. 
  The first error in the signal yield and branching fraction is 
  statistical, the second is systematic and the third is due to model dependence.
  The first error in the efficiency is due to MC statistics and systematic effects
  and the second one is due to model dependence.
}
 \begin{center}
 \begin{tabular}{lcccccc} \hline
  {Mode}              & Signal yield                               & Efficiency [\%]             & ${\cal{B}}$ [$\times10^{-7}$]    & U.L.[$\times10^{-7}$] &  Signif. \\ \hline
$K^0 e^+ e^-$         & $-1.0^{+1.8}_{-1.2}{}^{+0.2}_{-0.4}$  & $ 5.10\pm 0.29 \pm 0.20$ & $-0.70^{+1.29}_{-0.82}{}^{+0.14}_{-0.28} \pm 0.03$   & 3.0 & 0.0    \\ 
$K^+ e^+ e^-$         & $28.6^{+6.7}_{-6.0}{}^{+0.5}_{-0.7}$  & $ 16.3\pm 0.7 \pm 0.1$   & $ 6.40^{+1.50}_{-1.34}{}^{+0.29}_{-0.31}\pm 0.05 $             & -   & 6.4  \\ 
$Ke^+ e^-   $         & $26.6^{+6.8}_{-6.1}{}^{+0.6}_{-0.8}$  & $10.7\pm 0.5 \pm 0.1$    & $ 4.54^{+1.16}_{-1.04}{}^{+0.22}_{-0.24}\pm 0.06$             & -   & 5.5  \\ \hline
$K^{*0}e^+ e^-$       & $22.0^{+6.6}_{-5.9}{}^{+0.9}_{-0.7}$  & $ 4.32\pm 0.22 \pm 0.36$ & $18.5^{+5.5}_{-4.9}\pm1.1 \pm 1.5$   & -  & 4.4    \\
$K^{*+}e^+ e^-$       & $ 6.2^{+4.1}_{-3.4}{}^{+0.3}_{-0.6}$  & $ 1.42\pm 0.08 \pm 0.06$ & $16.0^{+10.4}_{-8.7} {}^{+1.2}_{-1.8} \pm 0.7$ & 37  & 1.7    \\ 
$K^* e^+ e^- $        & $28.9^{+7.6}_{-6.9} \pm 0.9        $  & $ 2.87\pm 0.15 \pm 0.21$ & $18.4^{+4.8}_{-4.4}\pm1.1 \pm 1.3$    & -   & 4.8  \\ \hline\hline
$K^0\mu^+\mu^-$       & $11.6^{+4.0}_{-3.4}{}^{+0.1}_{-0.3}$  & $ 6.76\pm 0.41 \pm 0.04$ & $ 6.26^{+2.17}_{-1.81}{}^{+0.38}_{-0.41} \pm 0.04$            & -   & 5.1  \\
$K^+\mu^+\mu^-$       & $39.9^{+7.6}_{-6.9}{}^{+0.5}_{-0.7}$  & $23.2\pm 1.1 \pm 0.5$    & $ 6.28^{+1.19}_{-1.08}{}^{+0.30}_{-0.31} \pm 0.13$            & -   & 8.3  \\ 
$K\mu^+\mu^-  $       & $51.5^{+8.4}_{-7.8}{}^{+0.5}_{-0.7}$  & $15.0\pm 0.7 \pm 0.2$    & $ 6.26^{+1.03}_{-0.64}{}^{+0.31}_{-0.32} \pm 0.10$            & -   & 9.7  \\ \hline
$K^{*0}\mu^+\mu^-$    & $40.7^{+7.6}_{-6.9}{}^{+0.5}_{-0.6}$  & $ 8.01\pm 0.43 \pm 0.14$ & $18.5^{+3.5}_{-3.1} \pm 1.0 \pm 0.3$            & -   & 8.4  \\
$K^{*+}\mu^+\mu^- $   & $11.4^{+4.5}_{-3.8}{}^{+0.2}_{-0.6}$  & $ 2.55\pm 0.14 \pm 0.08$ & $16.3^{+6.4}_{-5.4}{}^{+0.9}_{-1.2} \pm 0.5$            & -  & 3.5    \\ 
$K^*\mu^+\mu^-    $   & $52.5^{+8.7}_{-8.0}{}^{+0.5}_{-0.8}$  & $ 5.28\pm 0.29 \pm 0.03$ & $18.1^{+3.0}_{-2.8}\pm1.1 \pm 0.1$            & -   & 9.1  \\ \hline\hline
$K^0 \ell^+ \ell^-$   & $10.7^{+4.4}_{-3.7}{}^{+0.2}_{-0.5}$  & $ 5.95\pm 0.36 \pm 0.11$ & $ 3.28^{+1.34}_{-1.13}{}^{+0.21}_{-0.25} \pm 0.06$            & - & 3.5    \\ 
$K^+ \ell^+ \ell^-$   & $68.6^{+9.9}_{-9.3}{}^{+0.8}_{-1.0}$  & $ 19.8\pm 0.9 \pm 0.2$   & $ 6.32^{+0.92}_{-0.85} \pm 0.29 \pm 0.06$            & -   & 9.4  \\ 
$K \ell^+\ell^-   $   & $78.5^{+10.7}_{-10.0}{}^{+0.8}_{-1.1}$& $13.0\pm 0.6 \pm 0.4$    & $ 5.50^{+0.75}_{-0.70} \pm 0.27 \pm 0.02$            & -   & 11.0  \\ \hline
$K^{*0}\ell^+ \ell^-$ & $63.8^{+9.9}_{-9.2}\pm0.9          $  & $ 6.88\pm 0.37\pm 0.30$ & $16.9^{+2.6}_{-2.4} \pm 0.9 \pm 0.7$            & -    & 9.4   \\
$K^{*+}\ell^+ \ell^-$ & $16.3^{+5.6}_{-4.9}{}^{+0.5}_{-0.8}$  & $ 2.22\pm 0.13 \pm 0.02$ & $13.4^{+4.6}_{-4.0}{}^{+0.9}_{-1.0} \pm 0.1$    & -  & 3.7   \\ 
$K^* \ell^+\ell^-   $ & $82.2^{+11.4}_{-10.7}{}^{+1.0}_{-1.1}$& $ 4.55\pm 0.24 \pm 0.12$ & $16.5^{+2.3}_{-2.2} \pm 0.9 \pm 0.4$            & -   & 10.1    \\ \hline \hline
  \end{tabular}
\end{center}

\label{tab:results}
\end{table}

We consider experimental systematic effects from the fit, the efficiency 
determination and $B\overline{B}$ event counting.
Uncertainty in the background function is the dominant source 
of the systematic error.
To evaluate the change caused by the uncertainty in the signal function parameters,  
the mean and the width of the Gaussian
are changed by $\pm 1 \sigma$ from the values 
determined from $J/\psi K^{(*)}$ events.
The uncertainty in the background shape is obtained by varying 
the ARGUS shape parameter by $\pm 1 \sigma$ from the value determined 
with a large MC sample. The uncertainty in the peaking background
contribution is 
evaluated by changing the area of the associated Gaussian by $\pm 1 \sigma$.
The systematic errors associated with the fit function are shown 
in the second column of Table \ref{tab:results}.
Systematic uncertainties in the tracking, charged kaon identification, 
charged pion identification, electron identification, muon identification, $K^{0}_{S}$ detection 
and $\pi^{0}$ detection efficiencies are estimated to be 1.0\%, 
1.0\%, 0.8\%, 0.5\%, 1.2\%, 4.5\% and 2.7\% 
per particle, respectively.
The uncertainty in the background suppression is estimated to be 2.3\%
using $J/\psi K^{(*)}$ control samples. The systematic error due to MC 
statistics is less than 0.7\%.
The uncertainty in $B\overline{B}$ event counting is 1.1\%.
The systematic errors associated with efficiency and $B\overline{B}$ event
counting are summarized in Table \ref{tab:syst}.
Total experimental systematic errors are calculated by adding all systematic errors in quadrature.

The systematic uncertainty due to theoretical modeling is also evaluated.
We apply the same selection criteria on the signal samples generated
according to the three form factor models~\cite{ALGH,MNS,CFSS}
and obtain the efficiencies. The maximum 
difference in these efficiencies is assigned as uncertainty in model
dependence and is listed as the final value in column four of Table~\ref{tab:results}.

\begin{table}[htdp]
\begin{center}
\begin{tabular}{lccccc}\hline
Source                & $K^{0}$   & $K^{+}$ & $K^{+}\pi^{-}$ & $K^{0}_{S}\pi^{+}$ & $K^{+}\pi^{0}$  \\ \hline
Tracking              & 2.0       & 3.0     & 4.0            & 3.0                & 3.0             \\
Kaon identification               & -         & 1.0     & 1.0            & -                  & 1.0             \\
Pion identification               & -         & -       & 0.8            & 0.8                & -               \\
Lepton identification~($e$/$\mu$) & 1.0/2.4   & 1.0/2.4 & 1.0/2.4        & 1.0/2.4            & 1.0/2.4         \\
$K^{0}_{S}$ detection & 4.5       & -       & -              & 4.5                & -               \\
$\pi^{0}$ detection   & -         & -       & -              & -                  & 2.7             \\
BG suppression        & 2.3       & 2.3     & 2.3            & 2.3                & 2.3             \\
$B\overline{B}$ event counting & 1.1   & 1.1      & 1.1            & 1.1            & 1.1         \\ \hline
total~($e$/$\mu$)     & 5.8/6.1   & 4.2/4.7 & 5.1/5.5        & 6.7/6.8            & 6.3/6.1       \\ \hline
\end{tabular}
\end{center}
\caption{Summary of systematic errors in efficiencies and
$B\overline{B}$ event counting.}
\label{tab:syst}
\end{table}

In the calculation of  the branching fraction, we assume equal production rates
of charged and neutral $B$ meson pairs from $\Upsilon(4S)$ as well as isospin
invariance. When combining the $K^* e^+ e^-$ and $K^* \mu^+
\mu^-$ modes, we assume the ratio of branching fraction of $K^* \mu^+ \mu^-$ to that of $K^* e^+
e^-$ to be 0.75~\cite{ALGH}. The combined
branching fraction corresponds to the muon mode. The branching fractions
are found to be
\begin{center}
	${\cal B}(B \to K \ell^{+} \ell^{-}) = (5.50^{+0.75}_{-0.70} \pm 0.27 \pm  0.02) \times 10^{-7}$\\
	${\cal B}(B \to K^* \ell^{+} \ell^{-}) = (16.5^{+2.3}_{-2.2} \pm 0.9 \pm  0.4) \times 10^{-7}$,
\end{center}
where the first error is statistical, the second is systematic,
and the third is due to model dependence. 
These values are consistent with the SM
predictions~\cite{ALGH,MNS,CFSS,Exclusive}.
The branching fraction of $B \to K^* \ell^+ \ell^-$ is slightly larger
than our previous result~\cite{BelleKstarll}. Since we now use NNLO effective Wilson coefficients~\cite{ALGH},
which gives larger $C_7$ and smaller $C_9$ values 
than the older NLO calculation, the efficiency of $B \to K^* \ell^+ \ell^-$ 
is smaller by about 12\%. As a result, the branching fraction 
is now higher. (With the NLO effective Wilson coefficients, the branching 
fraction becomes $14.7 \times 10^{-7}$, which is consistent 
with our previous result.)
The branching fractions of the other decay modes are listed in
Table~\ref{tab:results}. 

For the modes with a significance of less than 3.0, we also set
90\% confidence level (C.L.) upper limits for the branching fractions.
The 90\% C.L. upper limit yield $N$ is defined as $\int^N_0 {\cal{L}}(n)dn = 0.9 \int^{\infty}_0 {\cal{L}}(n)dn$.
The function ${\cal{L}}(n)$ is the likelihood with signal yield $n$,
where each signal and background shape parameter is
modified up or down by its systematic error in the direction that
increases the signal yield. The upper limits for the branching
fraction are then calculated by 
reducing the efficiency by its systematic error.

We calculate the ratios of branching fractions to muon and electron modes. 
The ratio of branching fraction 
of $B \to K \mu^+ \mu^-$ to $B \to K e^+ e^-$ (${\cal{R}}_{K\ell\ell}$)is sensitive to neutral Higgs 
emission from the internal loop in the two Higgs doublet model with large
$\tan \beta$~\cite{AtwoodWang}. 
If the Higgs contribution is sizable, this ratio is greater than unity.
The corresponding ratio for $B \to K^* \ell^+ \ell^-$
(${\cal{R}}_{K^*\ell\ell}$) is sensitive to
the size of photon pole and is predicted to be about 0.75 in the SM. The
ratios are measured to be  
\begin{center}
  ${\cal{R}}_{K\ell\ell} = 1.38^{+0.39}_{-0.41}{}^{+0.06}_{-0.07}$ \\
  ${\cal{R}}_{K^*\ell\ell} = 0.98^{+0.30}_{-0.31}\pm0.08$,\\
\end{center}
which are consistent with the SM predictions.

\begin{figure}[htbp]
\begin{center}
  \includegraphics[scale=0.9]{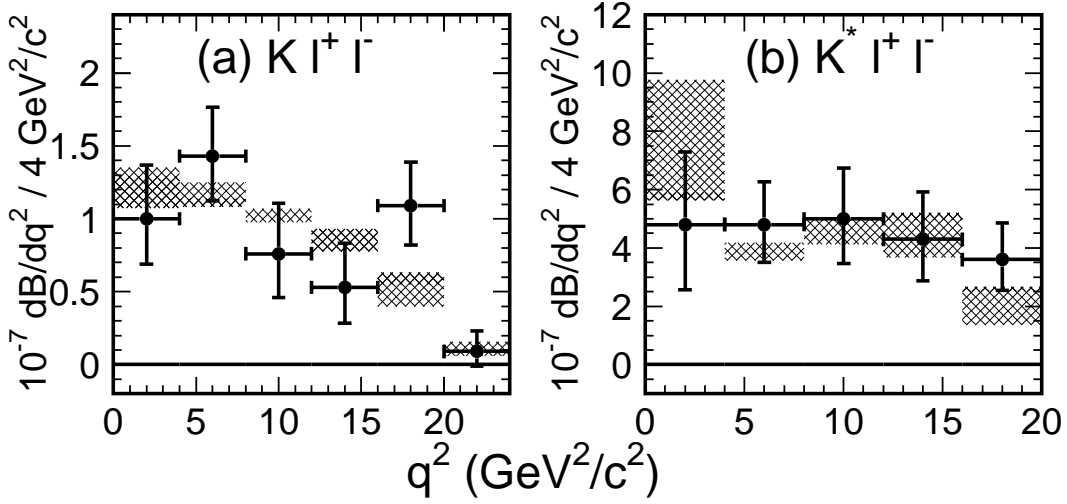}
\end{center}
\caption{
The $q^2$ distributions of  (a) $K \ell^+ \ell^-$ and (b) $K^* \ell^+
\ell^-$. Points with error bars show the data, while the hatched bands show
the range of the SM predictions~\cite{ALGH,MNS,CFSS}.
}\label{fig:q2}
\end{figure}

The distribution of the squared dilepton momenum $q^2$ and forward-backward asymmetries in 
$B \to K \ell^+ \ell^-$ and $K^* \ell^+ \ell^-$ are also measured. 
The forward-backward asymmetry ($A_{\rm FB}$) is defined as the partial rate 
asymmetry between the positive and negative regions of $\cos\theta_{B\ell^+}$,
the cosine of the angle between the $B^0$ or $B^+$ 
meson and positive charged lepton momentum directions in the dilepton rest frame,
\begin{center}
$\displaystyle A_{\rm FB}= \frac{\Gamma(\cos\theta_{B\ell^+} > 0) - \Gamma(\cos\theta_{B\ell^+} <0 )}{\Gamma(\cos\theta_{B\ell^+} > 0) + \Gamma(\cos\theta_{B\ell^+} <0 )}$.
\end{center}
The signal yield is extracted from a fit to the $M_{\rm{bc}}$
distributions in each $q^2$ bin and $\cos\theta_{B\ell^+}$ 
region. This procedure takes into account the forward-backward
asymmetry of the background. The $q^2$ resolution is about 0.6\% and
this is small enough relative to the bin size.
For the $q^2$ distribution measurement, the efficiency in each $q^2$ bin is 
obtained from the MC samples described earlier.
The branching fraction at low $q^2$ differs between $K^*e^+e^-$ and
$K^*\mu^+\mu^-$; we use the average in calculating the
$K^*\ell^+\ell^-$ differential branching fraction.
Figure~\ref{fig:q2} and Table~\ref{tab:q2} show the $q^2$ distributions; they
are in agreement with the SM predictions. 

\begin{figure}[htbp]
\begin{center}
  \includegraphics[scale=0.8]{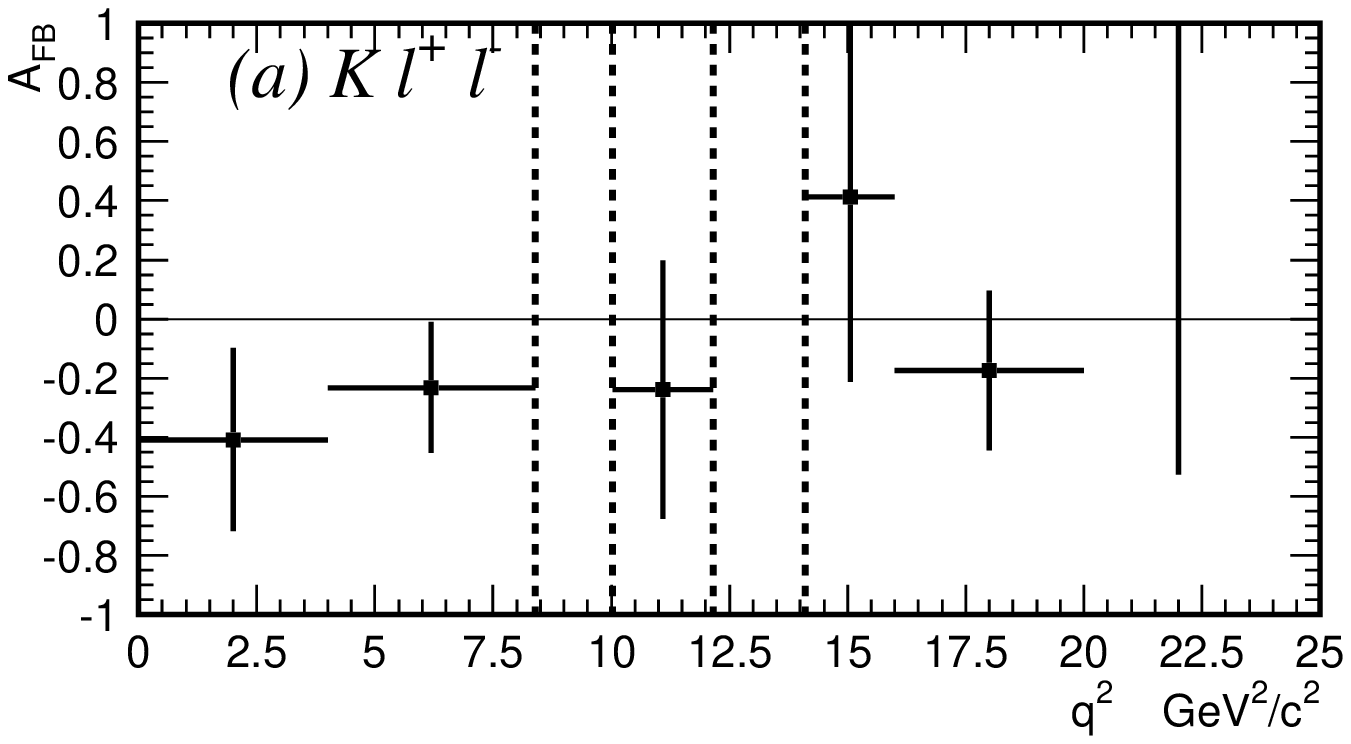}\\
  \includegraphics[scale=0.8]{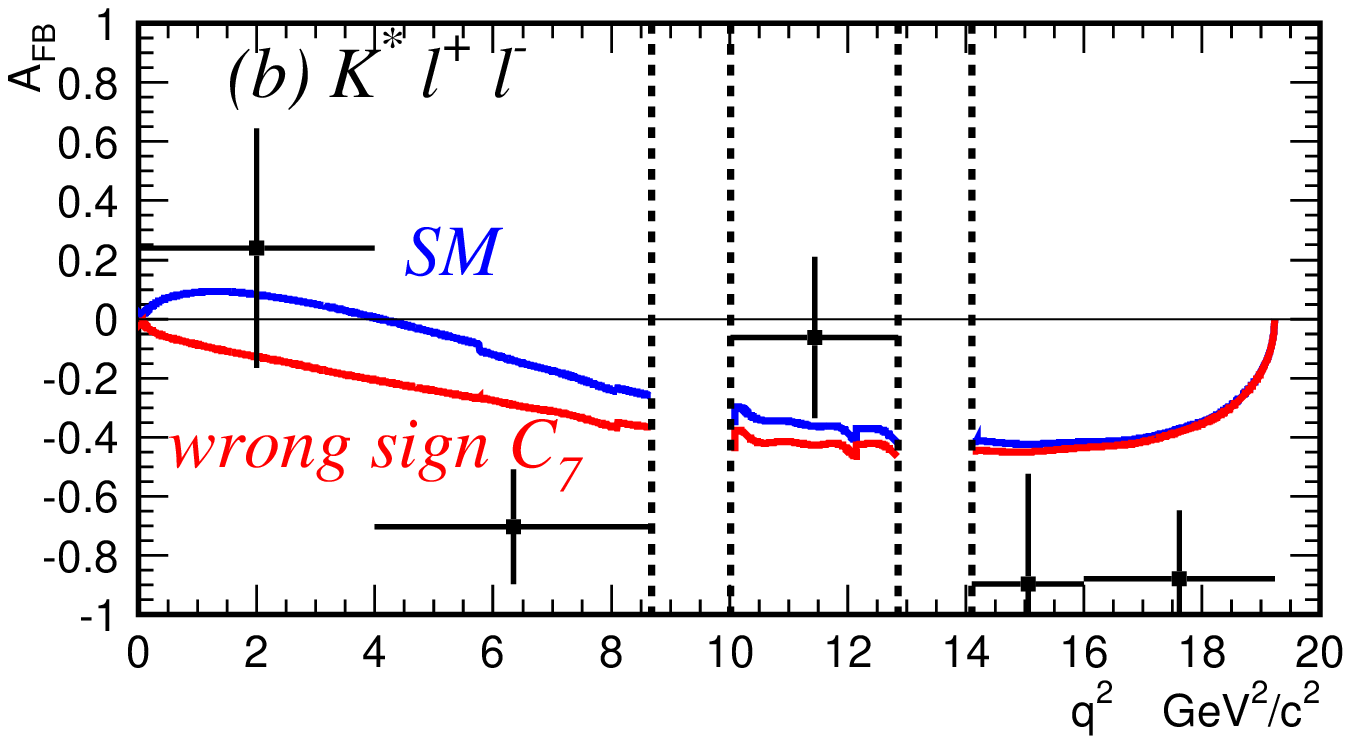}
\end{center}
\caption{
The raw forward-backward asymmetry in (a) $K \ell^+ \ell^-$ and (b) $K^* \ell^+
\ell^-$. Points with error bars show the data. Curves show
theoretical predictions including the effect of experimental
efficiency and resolution for
the SM case (blue) and wrong-sign $C_7$ case (red). The pairs of
dashed lines are the boundaries of the $J/\psi$ and $\psi'$ veto regions.
}\label{fig:afb}
\end{figure}

Figure~\ref{fig:afb} shows the raw forward-backward asymmetry distributions for 
$B \to K \ell^+ \ell^-$ and $K^* \ell^+ \ell^-$. The asymmetry in 
$B \to K \ell^+ \ell^-$ is expected to vanish in the SM; this
expectation is essentially unchanged by the presence of new
physics~\cite{NoAFBKll}. Therefore,
$B \to K \ell^+ \ell^-$ is a  
good control sample for the more interesting forward-backward
asymmetry in $B \to K^* \ell^+ \ell^-$. 
The measured asymmetry for $B \to K \ell^+ \ell^-$
is indeed consistent with zero. For $B \to K^* \ell^+ \ell^-$, the asymmetry
is in agreement with both the SM and the wrong-sign $C_7$ expectations,
since the statistical power is not yet sufficient to distinguish
between these two cases.

In summary, we report first measurement of forward-backward asymmetry
as a function of $q^2$ in $B \to K^* \ell^+ \ell^-$. Within the
limited statistical precision, the 
measured asymmetry in $B \to K^* \ell^+ \ell^-$ is consistent with
both the SM and the wrong sign $C_7$ case. 
We also report improved measurements of the branching fractions of $B
\to K \ell^+ \ell^-$ and $B \to K^* \ell^+ \ell^-$. The measured values, their
ratios and the $q^2$ distributions for $B \to K \ell^+ \ell^-$ and $B
\to K^* \ell^+ \ell^-$ are in good agreement with the SM
predictions~\cite{ALGH,MNS,CFSS,Exclusive}. 
\newpage

\begin{table}[htbp]
 \begin{center}
 \begin{tabular}{lccc}  \hline \hline
$K \ell^+ \ell^-$       &                 &                    &                               \\
$q^2/$GeV${}^2$   & yield                            & efficiency[\%]     & {\cal{B}}[$\times10^{-7}$] \\ \hline
$[0,4]$   & $12.4\PM{4.4}{3.7}\PM{0.3}{0.4}$ & $11.3\pm0.5\pm0.1$ & $1.00\PM{0.36}{0.30}\pm0.06\pm0.01$ \\ 
$[4,8]$   & $26.7\PM{6.1}{5.4}\pm 0.4$       & $17.0\pm0.8\pm0.3$ & $1.43\PM{0.32}{0.29}\pm0.07\pm0.01$ \\ 
$[8,12]$  & $10.1\PM{4.5}{3.8}\PM{0.4}{0.5}$ & $12.0\pm0.6\pm0.1$ & $0.76\PM{0.34}{0.29}\PM{0.05}{0.06}\pm0.01$ \\ 
$[12,16]$ & $ 6.5\PM{3.6}{2.9}\pm 0.3$       & $11.1\pm0.5\pm0.3$ & $0.53\PM{0.30}{0.24}\pm0.04\pm0.01$ \\ 
$[16,20]$ & $20.4\PM{5.4}{4.8}\PM{0.4}{0.5}$ & $17.0\pm0.8\pm0.3$ & $1.09\PM{0.29}{0.26}\PM{0.05}{0.06}\pm0.02$ \\ 
$[20,24]$ & $ 1.3\PM{2.1}{1.4}\pm 0.3$       & $13.2\pm0.6\pm0.8$ & $0.09\PM{0.14}{0.10}\pm0.02\pm0.006$ \\
 \end{tabular}

 \begin{tabular}{lccc}  \hline \hline
$K^* \ell^+ \ell^-$  &               &      &                                     \\ 
$q^2\GeV^2$   & yield                            & efficiency[\%]       & {\cal{B}}[$\times10^{-7}$] \\ \hline
$[0,4]$   & $11.3\PM{4.9}{4.2}\pm0.6$        & $2.16\pm0.11\pm0.56$ & $4.8\PM{2.1}{1.8}\pm0.4\pm1.2$ \\ 
$[4,8]$   & $21.6\PM{6.1}{5.4}\PM{0.4}{0.6}$ & $4.09\pm0.22\pm0.11$ & $4.8\PM{1.4}{1.2}\pm0.3\pm0.3$ \\ 
$[8,12]$  & $15.7\PM{5.3}{4.6}\pm0.4$        & $2.83\pm0.15\pm0.18$ & $5.0\PM{1.7}{1.5}\pm0.2\pm0.1$ \\ 
$[12,16]$ & $12.7\PM{4.7}{4.0}\pm 0.4$       & $2.71\pm0.14\pm0.07$ & $4.3\PM{1.6}{1.4}\pm0.2\pm0.1$ \\ 
$[16,20]$ & $16.0\PM{5.1}{4.4}\pm0.5$        & $3.99\pm0.21\pm0.26$ & $3.6\PM{1.2}{1.0}\pm0.2\pm0.2$ \\ \hline \hline
  \end{tabular}
\end{center}
\caption{Yield, efficiency and branching fraction in each $q^2$ region. The first error is statistical,
the second is systematic and the third is model dependence.}
\label{tab:q2}
\end{table}

   We thank the KEKB group for the excellent
   operation of the accelerator, the KEK Cryogenics
   group for the efficient operation of the solenoid,
   and the KEK computer group and the National Institute of Informatics
   for valuable computing and Super-SINET network support.
   We acknowledge support from the Ministry of Education,
   Culture, Sports, Science, and Technology of Japan
   and the Japan Society for the Promotion of Science;
   the Australian Research Council
   and the Australian Department of Education, Science and Training;
   the National Science Foundation of China under contract No.~10175071;
   the Department of Science and Technology of India;
   the BK21 program of the Ministry of Education of Korea
   and the CHEP SRC program of the Korea Science and Engineering 
Foundation;
   the Polish State Committee for Scientific Research
   under contract No.~2P03B 01324;
   the Ministry of Science and Technology of the Russian Federation;
   the Ministry of Education, Science and Sport of the Republic of 
Slovenia;
   the National Science Council and the Ministry of Education of Taiwan;
   and the U.S.\ Department of Energy.


\begin{thebibliography}{99}
\bibitem{cc}
Charge conjugate modes are implied throughout this paper.
\bibitem{aster}
Throughout this paper, variables denoted with an asterisk are
calculated in the $\Upsilon(4S)$ rest frame.
\end{thebibliography}
\end{document}